\documentclass[12pt, a4paper]{article}
\pdfoutput=1

\usepackage{amsmath}
\usepackage{amsfonts}
\usepackage{amssymb}
\usepackage{graphicx, rotating}
\usepackage{epstopdf}
\usepackage{epsfig}
\usepackage{latexsym}
\usepackage{graphicx}
\usepackage{color}
\usepackage{amsmath,amssymb}
\usepackage{cite}
\usepackage{slashed}
\usepackage{hyperref}

\numberwithin{equation}{section}

\hypersetup{colorlinks, citecolor=bluscuro, linkcolor=bluscuro, urlcolor=bluscuro}
\definecolor{rossos}{rgb}{0.8,0.2,0.3}
\definecolor{bluscuro}{rgb}{0.15, 0.2, 0.9}

\makeatother   

\newcommand{\bb}[1]{\textbf{#1}}
\newcommand{\vv}{ ``}

 \def\be   {\begin{equation}}   \def\ee   {\end{equation}}
 \def\ba   {\begin{array}}      \def\ea   {\end{array}}
 \def\bea  {\begin{eqnarray}}   \def\eea  {\end{eqnarray}}
 \def\bean {\begin{eqnarray*}}  \def\eean {\end{eqnarray*}}

\begin{document}


\vspace{0.5cm}
\begin{center}

\def\thefootnote{\fnsymbol{footnote}}

{\Large \bf Impact of the economic crisis\\
  \vskip 0.3cm on the Italian public healthcare expenditure}
\\[1.2cm]
{\large \textsc{Carlo Castellana}} 
\\[1cm]

{\small \textit{ Management School in Clinical Engineering,\\
University of Trieste, Italy}}

\vspace{.2cm}

\end{center}

\vspace{.8cm}

\begin{center}
\textbf{Abstract}
\end{center}

\noindent
The global financial crisis, beginning in 2008, took an historic toll on national economies around the world. Following equity market crashes, unemployment rates rose significantly in many countries: Italy was among those. 
What will be the impact of such large shocks on Italian \textit{healthcare} finances?
An empirical model for estimating the impact of the crisis on Italian public healthcare expenditure is presented. Based on data from epidemiological studies related to past economic crisis, the financial impact is estimated to be comparable to the healthcare deficit of Italian Regions (EUR 3-5 bn). 
According to current agreements between the Italian State and its Regions, public funding of regional National Health Services (NHSs) is limited to the amount of regional deficit and is subject to previous assessment of strict adherence to constraint on regional healthcare balance-sheet. Those Regions that will fail to comply to balance-sheet constraints will suffer cuts on their public NHS financing with foreseeable bad consequences for the health of their regional population.
The current crisis could be a good timing for a large-scale re-engineering of the Italian NHS, probably the only way for self-sustainability of the public system.

\def\thefootnote{\arabic{footnote}}
\setcounter{footnote}{0}
\pagestyle{empty}

\newpage
\pagestyle{plain}
\setcounter{page}{1}

\section{Introduction} \label{Introduction}
\noindent
Following the tech bubble and the events of September 11, the Federal Reserve stimulated a struggling economy by cutting interest rates to historically low levels. As a result, a housing bull market was created. In turn, investors sought higher returns through riskier investments. Lenders took on greater risks and approved subprime mortgage loans to borrowers with poor credit. Consumer demand drove the housing bubble to all-time highs. Interest rates climbed back up which caused default on many subprime mortgages. This left mortgage lenders with property that was worth less than the loan value due to a weakening housing market. Defaults increased, the problem snowballed and several lenders went bankrupt. 


The crisis was named \vv Subprime Crisis", which was fine at first, but it's now totally inadequate. 
It started as a phenomenon affecting one country (the USA) and one market sub-sector (the subprime mortgage market) only.
In few months the contagion widespread to the overall financial system where it took many facets: turmoil reached the global mortgage markets and later on credit markets where it became clear that this was not just a painless \vv infection" but a true \vv epidemic". A bank crisis ensued, thus hitting the heart of the overall financial system: the lack of liquidity forced exceptional interventions of the European Central Bank and Federal Reserve. 

By leaking into the real economic tissue, the crisis entered the real life of common people. The effects of the economic crisis in the USA are currently being felt worldwide and Italy is not an exception.

The crisis will likely have a long-lasting impact on Italy's economic potential \cite{IMFCountryReport}.
Indeed, innovation and investment opportunities may weaken because demand prospects are likely to be poor and the real cost of borrowing remains high. In addition, some of the increase in unemployment may be structural given that very difficultly displaced workers will be able to return to the labor market as industrial restructuring takes hold.

Italy's structural problems turned in a poor productivity performance and a reduced growth over the last fifteen years \cite{Fachin}. The global financial crisis has exacerbated these long-standing weaknesses, taking a heavy toll on Italy's economy. Italy has suffered from chronically low economic growth, even before the global financial crisis. Real Gross Domestic Product (GDP) growth averaged 1.6\% during the period 1995-2007, down from over 2\% in the earlier decade \footnote{Eurostat, data publicly available at \textit{http://epp.eurostat.ec.europa.eu}.}.
In the first quarter of 2009, growth witnessed a decline four times as large as the one experienced during the European Monetary System (EMS) crisis of 1992-93 \cite{Intro_Bassanetti}. 

While most forecasters were expecting the beginnings of recovery in late 2009 or early 2010, and even if many financial indicators point for a gradual general improvement in the main economies, the current situation has not recovered yet to the pre-crisis one and the actual duration will depend on the effectiveness of policies in the developed countries.\\

Financial markets crashes, the economy shrinks, people loose their jobs: how can this affect the Italian NHS?

\begin{figure}[h!]
\centering
  \includegraphics[scale=0.5]{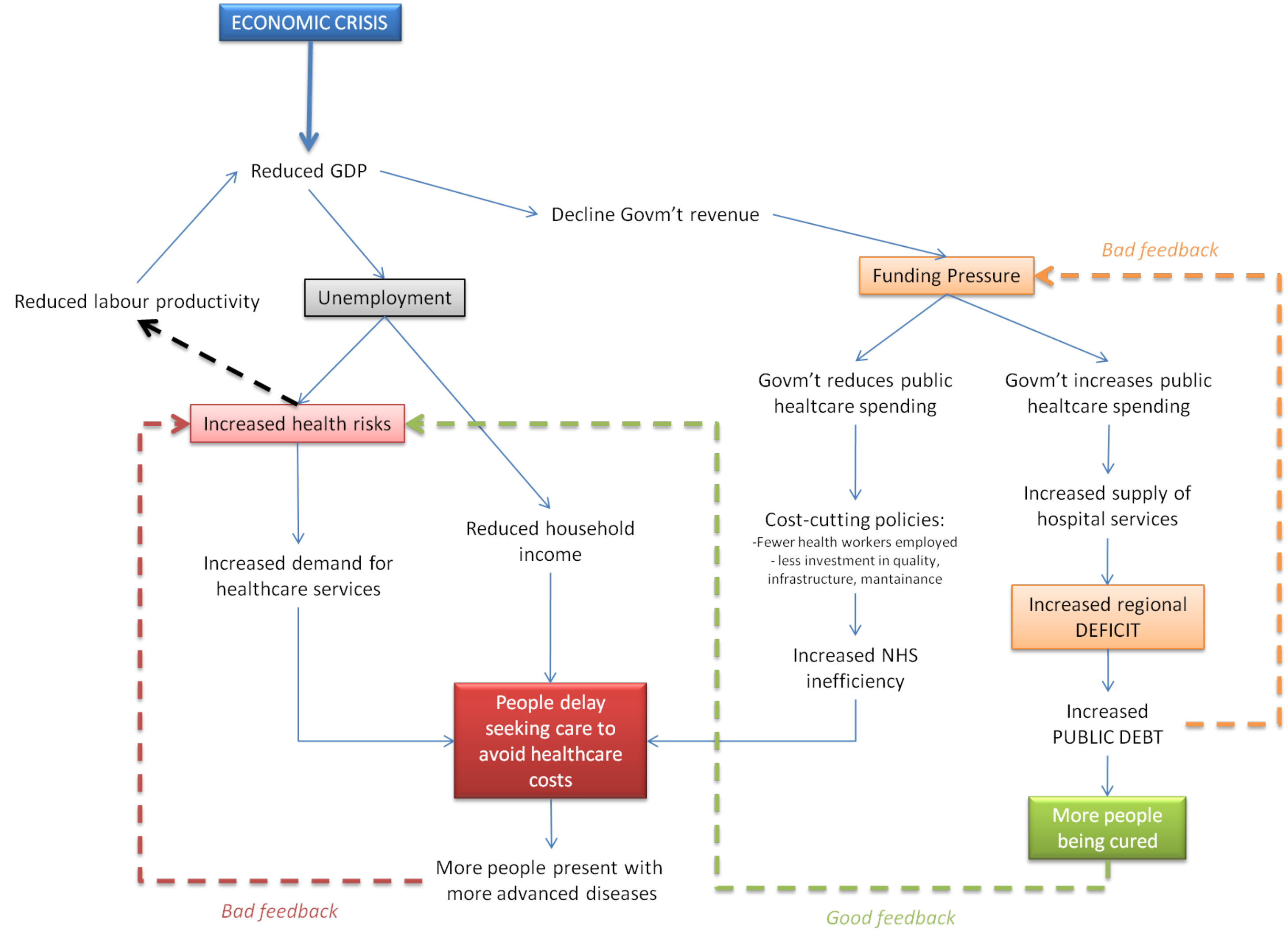}
\caption{The economic crisis induces a waterfall of events in the economic, healthcare, social and Government sectors. The main driver of fiscal policies is the funding pressure on Government balance sheet. Two opposite actions are possible: either increase or decrease healthcare spending. The two have different consequences that feed-back to Government funding pressure by three very different routes (\textit{Bad feedback 1, Good feedback 2 and Bad feedback 3 lines}).}
\label{fig: Intro_3}
\end{figure}

\begin{figure}[h!]
\centering
\includegraphics[scale=0.5]{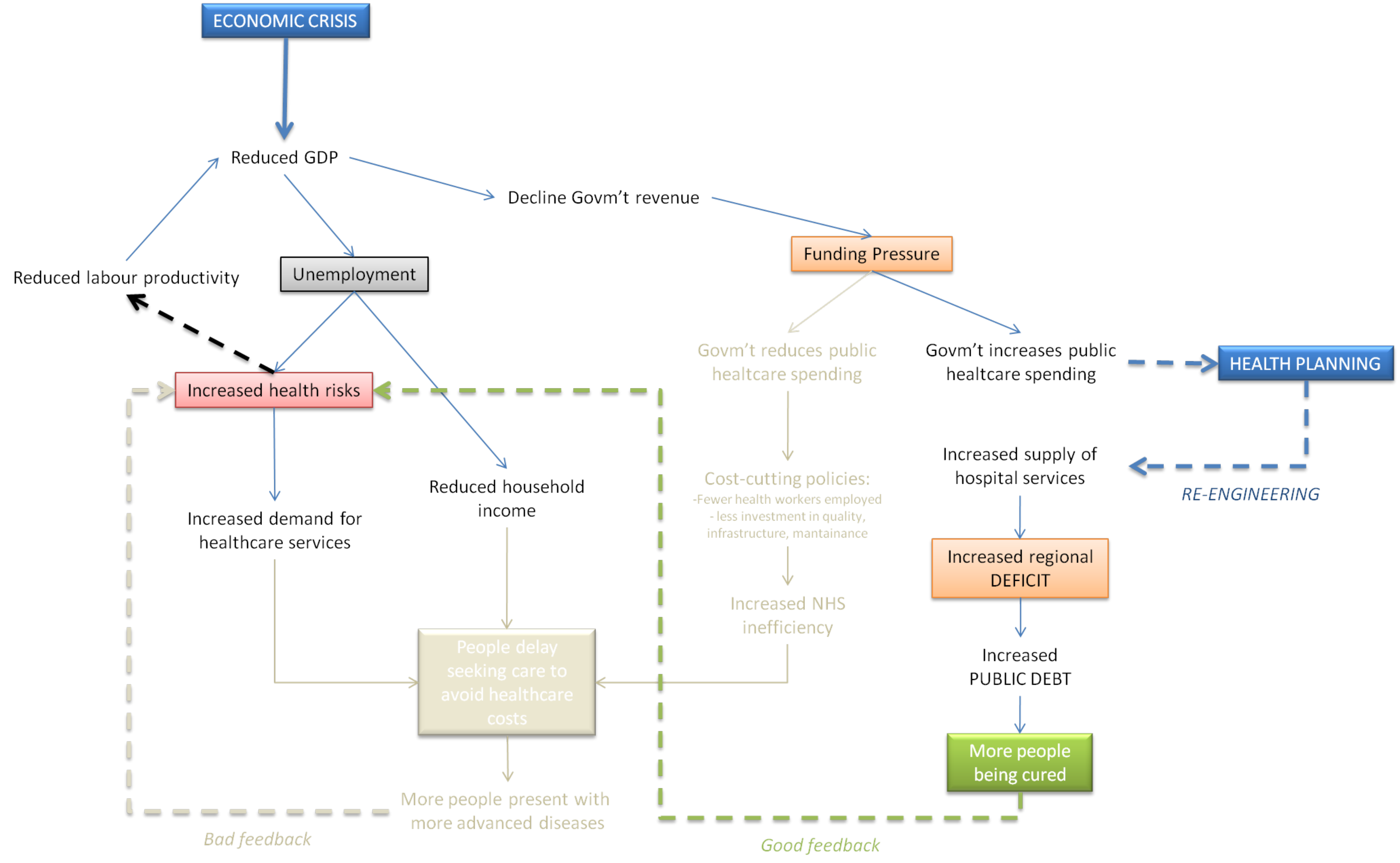}
\caption{Relationship between Health Planning and Economic crisis. Through a re-engineering of the public NHS (\textit{RE-ENGINEERING line}), Health Planning avoids increased funding pressures allowing the \textit{Good feedback 2} route to positively feed through health improvement and increased labour productivity. Compare with fig. \ref{fig: Intro_3}.}
\label{fig: Intro_4}
\end{figure}

\noindent The relation (fig. \ref{fig: Intro_3} and \ref{fig: Intro_4}) is that job losses mean a reduction of household income and this has serious impacts on people's health \cite{HealthEconStatus1, HealthEconStatus2, HealthEconStatus3, HealthEconStatus4,HealthEconStatus5,HealthEconStatus6,HealthEconStatus7,HealthEconStatus8}. 

\noindent Past economic crisis were associated with an increase in morbidity and mortality \cite{PastCrisisConseq1}, and this one is not going to be an exception \cite{SubpCrisisConseq1,SubpCrisisConseq2,SubpCrisisConseq3}. Understanding the routes of potential impact on health and healthcare services of the reduction of income for Italian individuals as a consequence of the crisis, provides the framework for action to mitigate against this threat becoming a reality.\\

\section{Methods}
\sectionmark{Forecasting the impact of the economic ..}
\label{sec:Methods}
%
\noindent 
The healthcare implications of the economic crisis can be numerous and serious. Before choosing how to act in order to cope for such bad consequences, one need to quantify their financial impact: any health plan to re-engineer the NHS will depend on the availability of enough financial resources to cover that plan.\\ 

\noindent \bb{What will be the impact of the current economic crisis on the Italian public NHS expenditure?}\\

\noindent To answer this question a financial model has been implemented. The model has three types of inputs \cite{Polder}:
\begin{enumerate}
\item 	\bb{demographics}: population, GDP, unemployment rate, mortality rates, life-expectancy;
\item 	\bb{healthcare}: healthcare expenditure, ratio between the healthcare expenditure of a person during its last year of life and that of a person with the same age but in good health conditions (\textit{Deceased/Survivor Ratio});
\item 	\bb{econometrics}: hypothesis on the \textit{main driver} of healthcare expenditure (GDP, healthcare personnel, \textit{death-related costs}).
\end{enumerate}

\noindent Predicting the future evolution of healthcare expenditure is one of the crucial challenges that the EU and its Member States are facing in the context of the demographic and social changes that are currently taking place in Europe. It is a hard and yet unknown subject which is constantly threatening the main economic panels worldwide \cite{EP1,EP2,EP3} due to the complexity of the systems and multiplicity of factors affecting both total and public spending. To tackle this issue a major project was undertaken by the European Commission and Economic Policy Committee which aimed at projecting future public healthcare expenditure in twenty-seven Member States of the EU and Norway over the period 2007- 2060 \cite{RE_Przywara}.

Two types of uncertainties make projecting the future evolution of the \textit{absolute} value of healthcare expenditure a very complex issue: one at model level (the multifactorial nature of healthcare spending) and another at the level of model parameters (the intrinsic volatility of variables underlying the model).

However, within this paper we are only interested in calculating the impact of the economic crisis on the healthcare expenditure which, thanks to its \textit{differential} nature \footnote{See eq. \ref{eq_CRIMI} and eq. \ref{eq_CRIUI}.}, has a lower level of uncertainty. 
The basic idea is that \vv noise" affecting two estimates of a variable calculated with the same model would net out when calculating their difference. As a consequence, the differential estimate will be much more reliable than the absolute estimates: within this level of approximation, the details of the underlying econometric model (EM) become less relevant.
Of course this is even truer if the differential quantity we are calculating is a \vv small" one compared to the other two entering the difference. The impact of the economic crisis on Italian healthcare expenditure falls within these assumptions: as it will be shown throughout this paper, its estimate is EUR 0.5-9.5 bn, far less than the total Italian NHS expenditure of EUR 109.7 bn \footnote{Source: Italian Ministry of Health, 2009.}.

Remarkably the actual financial impact of the crisis, even if being a small portion of the total healthcare expenditure, it is not negligible compared to the EUR 2.368 bn \footnote{Source: Italian Ministry of Health, 2009.} of \textit{healthcare deficit}. 
The knowledge of such an estimate becomes very important for those Institutions in charge of regional Health Planning, especially during the current difficult conditions of the Italian NHS that finds many Regions enforced to implement recovery plans, {\bf \vv Piani di Rientro"}, aimed at reducing the issue of the inflating regional healthcare deficit.\\

\section{Epidemiological analysis}
\label{subs: prevstu}
\noindent
How can we measure what the current economic crisis will mean for the health of Italian population? How much distress will the impact on health cause to the already fragile financial \vv health" of Italian regional balances?

\noindent We will answer by reviewing the epidemiological evidence supporting an association between unemployment and health outcomes in the population. 

\noindent In order to find some parameters that can help \vv measuring" the impact of the crisis on healthcare, we review the experience of two major economic crises of the 20th century in addition to various epidemiological studies on specific cohorts:

\begin{enumerate}
\item 	Post-communist Depression (early 1990s) \cite{Impact_Men} and \cite{Impact_Sachs, Impact_Wedel, Impact_Murrell, Impact_Kontorovich}; 
\item 	East Asian crisis (late 1990s) \cite{Impact_Kim, Impact_Khang, Impact_Waters} and \cite{Moreno, Intereconomics};
\item 	studies on specific cohorts:
	\begin{enumerate}
	\item 	Canadian recession (1985) \cite{Impact_Siddique}
	\item 	British cohorts (1981-1983 and 1971-1981) \cite{Impact_Moser1, Impact_Moser2, Impact_Beale1, Impact_Beale2, Impact_Yuen};
	\item  	Finnish cohort (1981-1985) \cite{Impact_Martikainen};
	\item  	Danish cohort (1970-1980) \cite{Impact_Iversen};
	\item 	Italian cohorts (1976-1985 and 1981-1985) \cite{Impact_Costa} ;
	\item  	US veterans cohort \cite{Impact_Linn}
 	\end{enumerate}
\end{enumerate}

\noindent Available evidence suggests that health is at risk in times of rapid economic change, in both booms and busts. However the impact on mortality is exacerbated where people have easy access to the means to harm themselves and is ameliorated by the presence of strong social cohesion and social protection systems.

\noindent We can look to experiences of the past to guide our expectations of the public health effects of this crisis. 

\noindent Reported results of studies on specific cohorts are based on the review of \cite{Impact_Jin} where the authors selected articles published in 1980s and 1990s supporting an association between unemployment and adverse health outcomes.
Those include time-series analysis, cross-sectional, case-control and cohort studies.

\subsubsection*{Approximation n.1} \label{sec: Approx1}
We translate the effect of the crisis into a perturbation of some parameters entering our model of public healthcare expenditure. As a first approximation the impact of the crisis on healthcare has been assumed to depend on (see fig. \ref{fig: ReE_1}):
\begin{enumerate}
\item 	mortality rates (demographic measure)
\item 	utilization rate of healthcare services (healthcare measure)
\end{enumerate}


\subsubsection*{Approximation n.2}
The second type of approximation has been used to address the way the increased death relative risk (RR) translates into increased mortality rates. Independently of the term-structure of mortality rates, we adopt the following simplifying relationship:
\begin{equation} \label{eq: Impact_1}
 PD' = \frac{UEP \cdot PD \cdot RR + EP \cdot PD}{WAP}
\end{equation}
where PD and PD' are the probability of death in 5 years time \footnote{Based on previous studies  \cite{Impact_Kim, Impact_Jin}, the impact of other economic crisis on mortality has been reported to last from 2 to 10 years. We consider a 5 years period as a reasonable \vv average" duration.} that should be expected under normal circumstances and, respectively, as a consequence of the economic crisis; \textit{UEP} and \textit{EP} are the number of people that are unemployed and, respectively, employed; \textit{WAP} is the active population, i.e. the number of people in working-age cohorts.


\noindent Eq. \ref{eq: Impact_1} can be rearranged as:
\begin{equation} \label{eq: Impact_2}
 PD' = PD \cdot \frac{UEP \cdot RR + EP }{WAP}
\end{equation}
that, together with \textit{WAP}=\textit{UEP} + \textit{EP}, yields:
\begin{equation} \label{eq: Impact_3}
 PD' = PD \times RR'
\end{equation}
and based on the definition of unemployment rate $\omega$, the effective RR can be calculated as:
\begin{equation} \label{eq: Impact_4}
RR' = 1 + \frac{UEP}{WAP} \cdot (RR-1)  = 1 + \omega \cdot (RR-1)
\end{equation}

\vskip 5mm

\noindent Based on what stated so far we can now answer to the following two questions.\\

\begin{figure}[h!]
\centering
\includegraphics[scale=0.5]{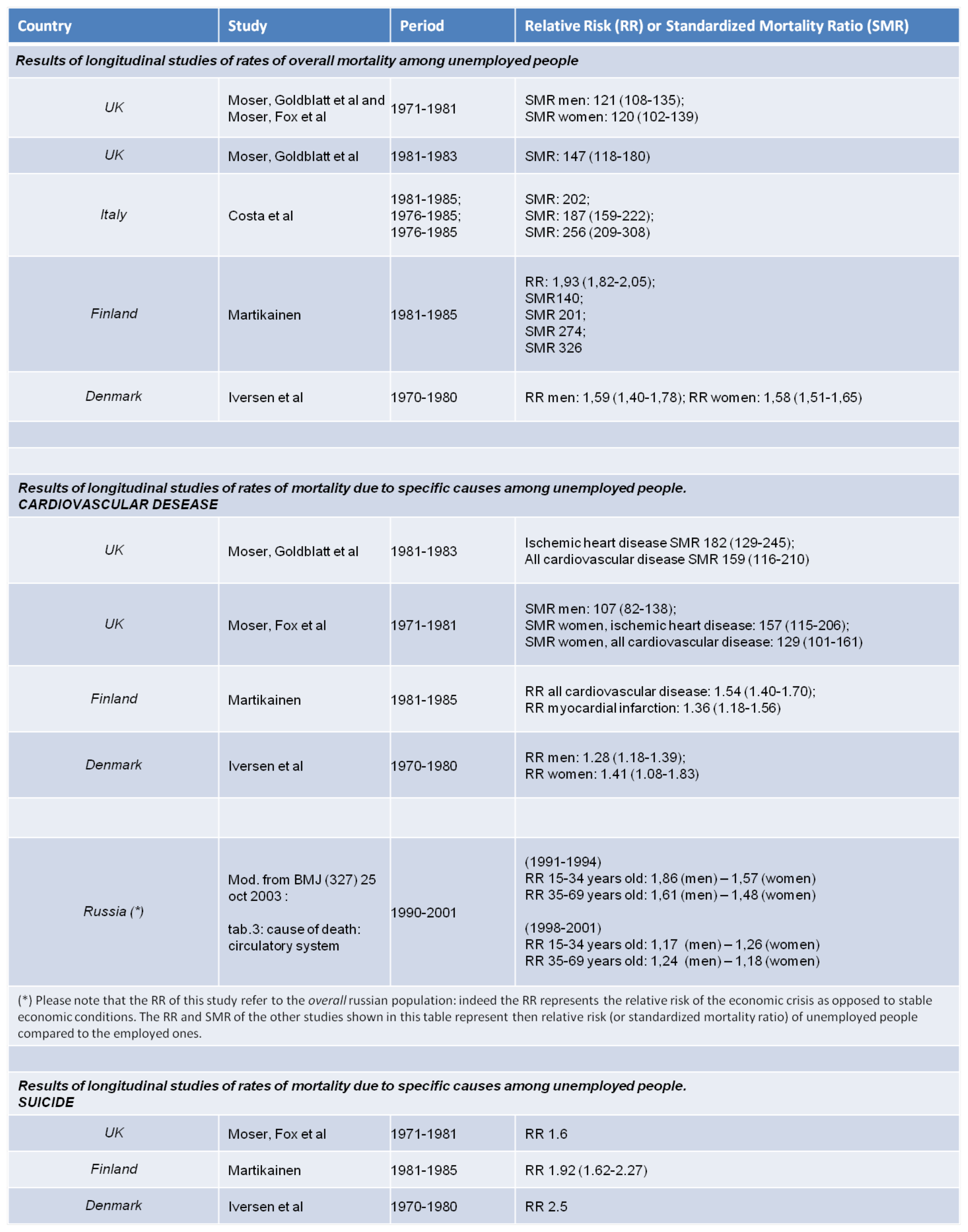}
\caption{Results of longitudinal studies of rates of overall mortality among unemployed people due to general and specific causes. Please note that SMR are expressed in hundreds, i.e. to compare SMR with RR one should firstly divide the shown SMR number by 100. Numbers in brackets are 95\% confidence range. \textit{RR}: relative risk. \textit{SMR}: standardized mortality ratio. \textit{Source: mod. from \cite{Impact_Jin}.}}
\label{fig: Impact_3}
\end{figure}

\noindent \bb{Q.1 } \textit{What number will we use to quantify the mortality increase due to the economic crisis?}
\\ As a first step, we need to make sure data on which we extrapolate the mortality increase to be homogeneous.
As explained in fig. \ref{fig: Impact_3} the Russian data are diluted to the overall population and hence are not directly comparable to the \vv undiluted" ones of the other studies.\\
By means of eq. \ref{eq: Impact_4}, assuming a 10\% current level of unemployment rate \footnote{According to Eurostat, 2011Q4 unemployment rate amounts to 9.1\% in Italy and to 10.6\% in Europe (17 Countries). \href{http://epp.eurostat.ec.europa.eu/portal/page/portal/statistics/search_database}{[Eurostat]}}, we are able to compare RRs values of cohort studies that applied to the subpopulation of unemployeds only (data in fig. \ref{fig: Impact_3}), to RRs values reported for the Russian crisis. 

\noindent Results of this process are reported in fig. \ref{fig: Impact_5} where the selected range for each age-cohort is shown.\\


\vskip 2mm

\begin{figure}[h!]
\centering
\includegraphics[scale=0.5]{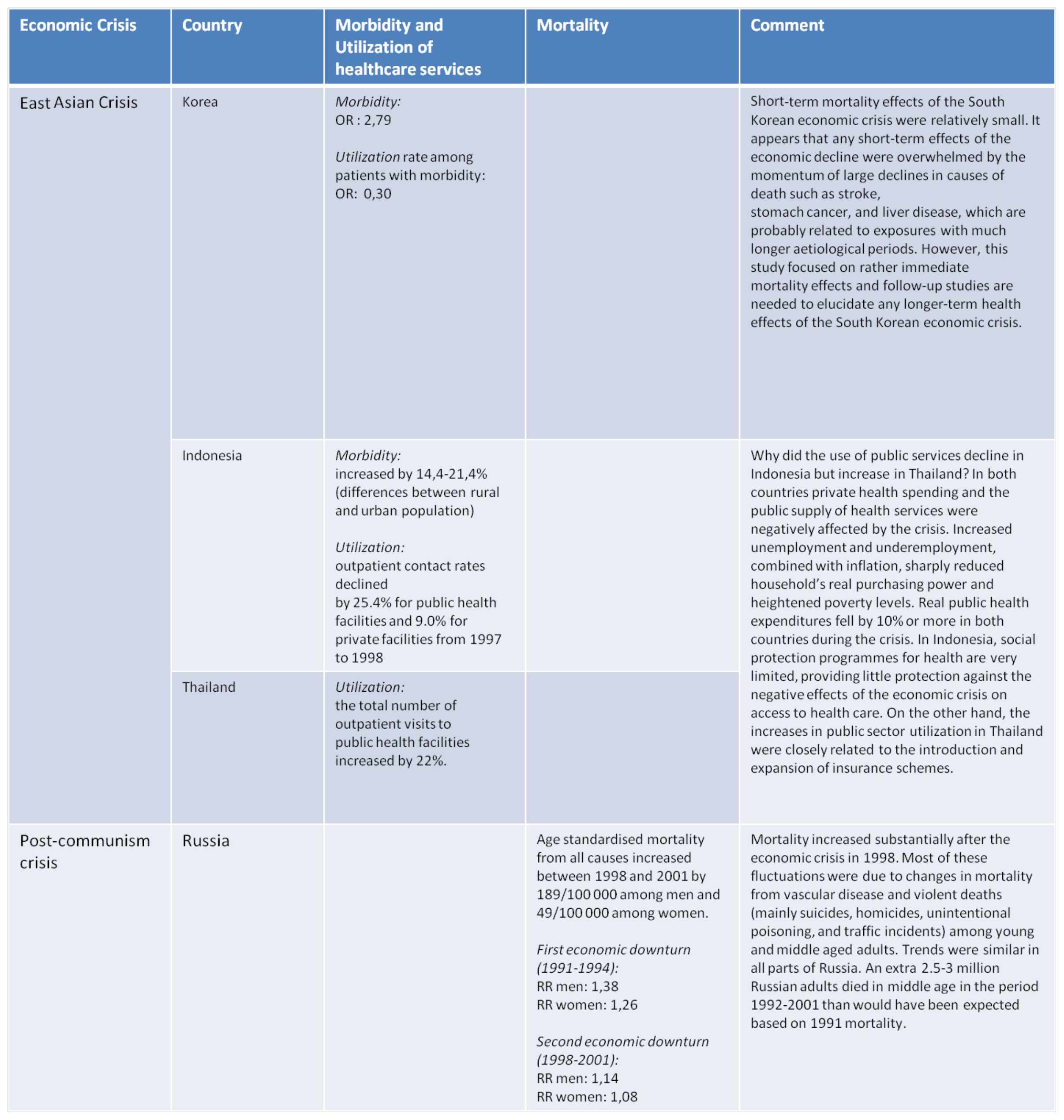}
\caption{Summary of the main measured health impacts of the two late 20th century's crisis. Numbers in brackets are 95\% confidence range. \textit{RR:} relative risk. \textit{OR:} odds ratio. \textit{Source: \cite{Impact_Men, Impact_Kim, Impact_Khang, Impact_Waters}}  }
\label{fig: Impact_2}
\end{figure}

\begin{figure}[h!]
\centering
\includegraphics[scale=0.5]{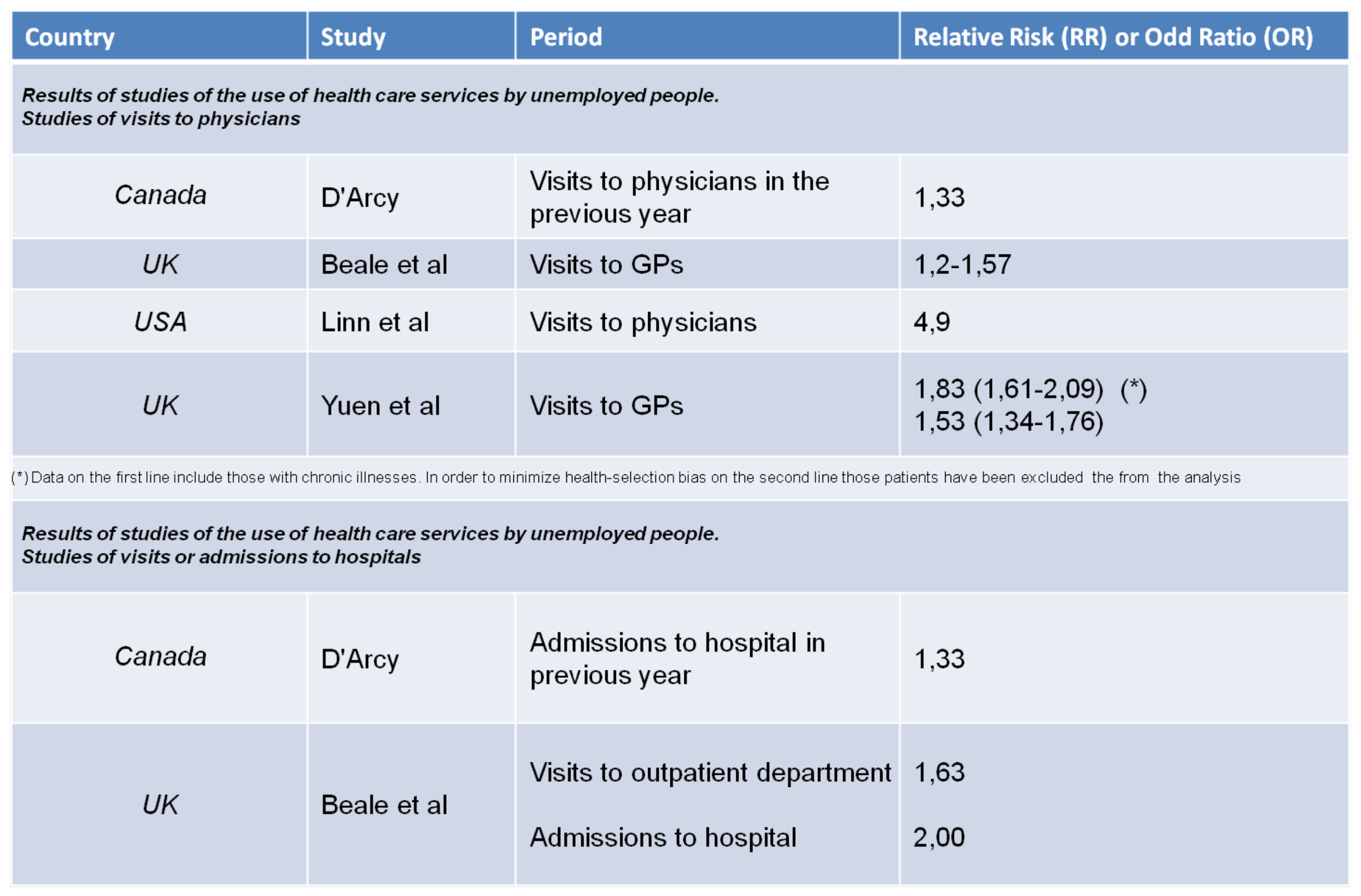}
\caption{Results of studies of the use of healthcare services by unemployed people. Studies of visits to physicians (top) and of visits or admissions to hospitals (bottom). Numbers in brackets are 95\% confidence range. \textit{Source: mod. from \cite{Impact_Jin}}.}
\label{fig: Impact_4}
\end{figure}

\noindent \bb{Q.2 } \textit{What number will we use to quantify the increased usage of healthcare services due to the economic crisis?}
\\The use of healthcare services is strongly dependent on the type of healthcare system. 
Indeed the association of unemployment with increased use of healthcare services depends on services being universally available and free of charge at point of use, as they are in Canada and Britain. In the USA, in contrast, hard economic times may mean \vv nearly empty waiting rooms" because jobless people often lack health insurance and the ability to pay \cite{Impact_Frey}. 
The same applies to the East Asia: as reported in fig. \ref{fig: Impact_2}, the effect of the East Asian crisis on countries as Korea and Indonesia was a decline in usage of healthcare services. This is not surprising given that Korean healthcare system can be described as a privately controlled delivery system in combination with a publicly regulated financing system \cite{Impact_Kim} and, as of 1998, health insurance was an income-based contributory insurance program. Indonesia was lacking of social protection schemes. \\
The Italian NHS is much more similar to the UK, Swedish and Canadian ones than it is to the US or East Asian ones. 
Due to this important difference averaging numbers of cohort studies - which, as reported in fig. \ref{fig: Impact_4}, refer to Canada, UK and US veterans - with those coming from studies on East Asian crisis can turn into meaningless results. 

Henceforth we will discard data of fig. \ref{fig: Impact_2} and we will look at data in fig. \ref{fig: Impact_4} only. 
Several studies \cite{Impact_Siddique, Impact_Beale1, Impact_Beale2, Impact_Linn, Impact_Yuen} reported increased use of general healthcare services in association to an increased unemployment rate. These studies report increases in visits to physicians, hospital inpatient or outpatient admissions and use of prescription medication.

Limited by the available statistics reported in the aforementioned studies, we selected the following ranges to quantify the usage of healthcare services:
\begin{itemize}
\item we take 1.20 - 1.57 as range for the increase in usage of general practitioners (GP) services \footnote{Note: data on the first line of the Yuen et al. study in fig. \ref{fig: Impact_4} include those with chronic illnesses. In order to minimize health-selection bias we only accounted for the value 1.53 reported on the second line.};
\item for the increase in hospital utilization rates we use the only two values available: 1.33 - 2.00;
\item for the Outpatient visits we use the only data available: 1.63.
\end{itemize}
We translate the above ranges according to eq. \ref{eq: Impact_4} and get: 1.02 - 1.057 for the Visits to GPs, 1.033 - 1.10 for Admissions to hospitals and 1.063 for the Outpatient visits (see fig. \ref{fig: ReE_1}). 

\vskip 5mm

\section{The financial model}
\label{subs: finmod}
%
\noindent
Given the wide range of underlying factors and channels through which they affect spending, rather than attempting to construct an all-encompassing projection methodology to capture all demographic and non-demographic factors, two projection scenarios have been run in order to tackle the issue from different perspectives:
\begin{enumerate}
	\item pure demographic (PD);
	\item constant health (CH);
\end{enumerate}

\noindent A discussion of the mathematical details of each one of those econometric projection scenarios can be found in \cite{RE_Przywara}.\\ 

\begin{figure}[h!]
\centering
\includegraphics[scale=0.6]{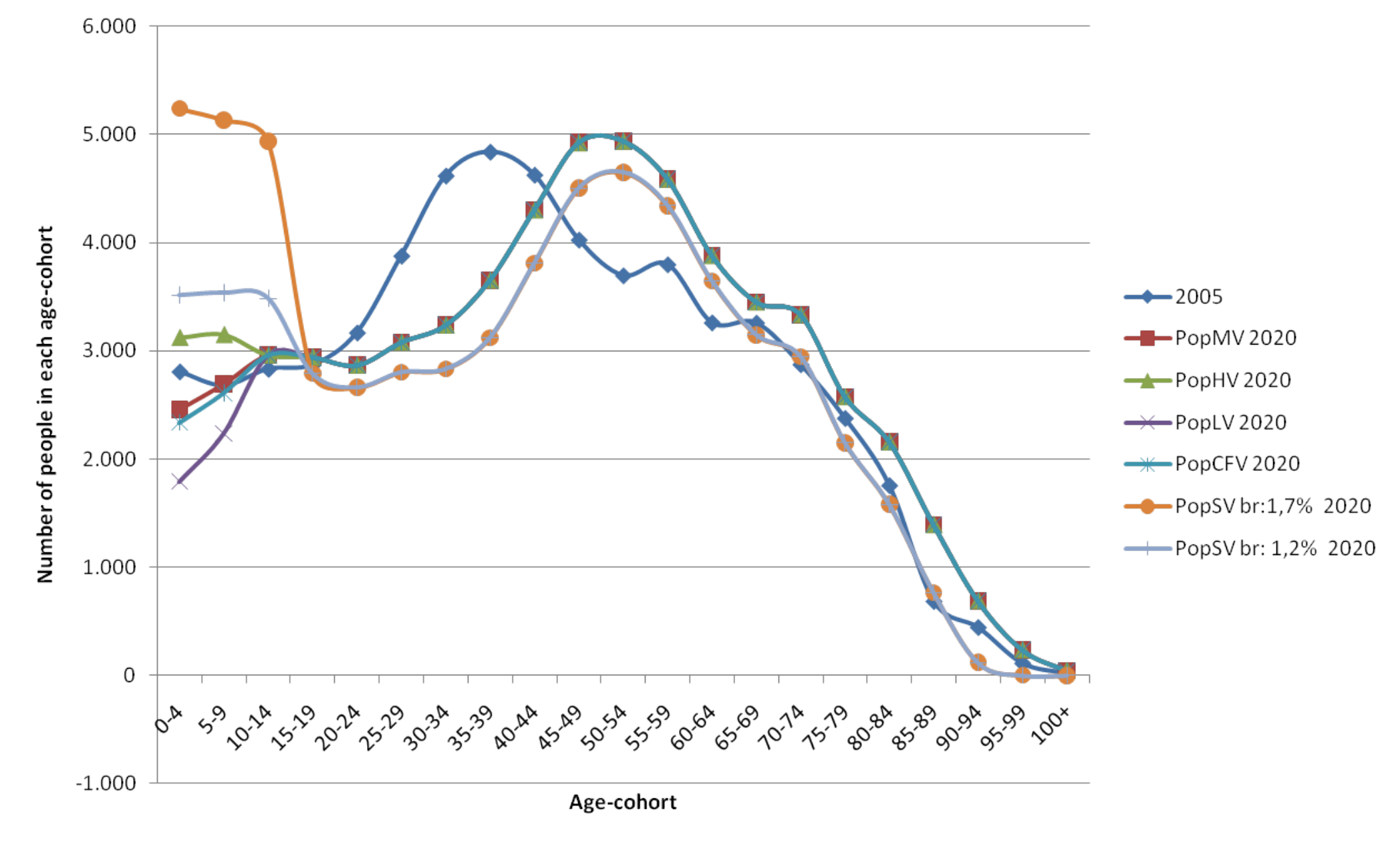}
\caption{Population projections. Four scenarios (PopMV, PopHV, PopLV, PopCFV) are shown based on different assumptions on mortality and birth rates. Value shown are in thousands. PopSV br: 1.7\% (resp. PopSV br: 1.3\%) is the HIGH (resp. LOW) birth-rate scenario. \textit{Source: http://esa.un.org/unpp}.}
\label{fig: ReE_4}
\end{figure}

\noindent Population projections have been modeled according to the six scenarios of fig. \ref{fig: ReE_4}:
\begin{enumerate}
\item four scenarios (PopMV, PopHV, PopLV and PopCFV) are available on the European Statistical Agency (ESA) website and are based on ESA assumptions. Indeed, the four scenarios differ by the distributions of persons among age-cohorts, PopLV representing a relatively \vv older" population, PopHV a \vv younger" one and PopMV an \vv average" between the two. PopCFV is a \vv constant fertility variant". Please note that these four scenarios are projected independently of the mortality rates scenarios \footnote{However, as it will be demonstrated hereafter, this is not a big issue: being interested in a differential effect, the detailed assumptions on the future projections of the different variables (population, D/S ratio, etc.) would affect the result as a second order correction (i.e. for what we are concerned in this study, they are negligible).};
\item two additional scenarios (PopSV br: 1.7\% and PopSV br: 1.3\%) have been simulated starting from current mortality rates \footnote{As opposed to the other four scenario, this time we are making sure consistency between population and mortality rate projections is preserved.}. These two scenarios are calculated based on current population structure by age-cohorts, mortality tables and a crude birth rate of 1.7\% (for the HIGH birth-rate scenario) and 1.3\% (for the LOW birth-rate scenario).
\end{enumerate}

\begin{figure}[h!]
\centering
\includegraphics[scale=0.6]{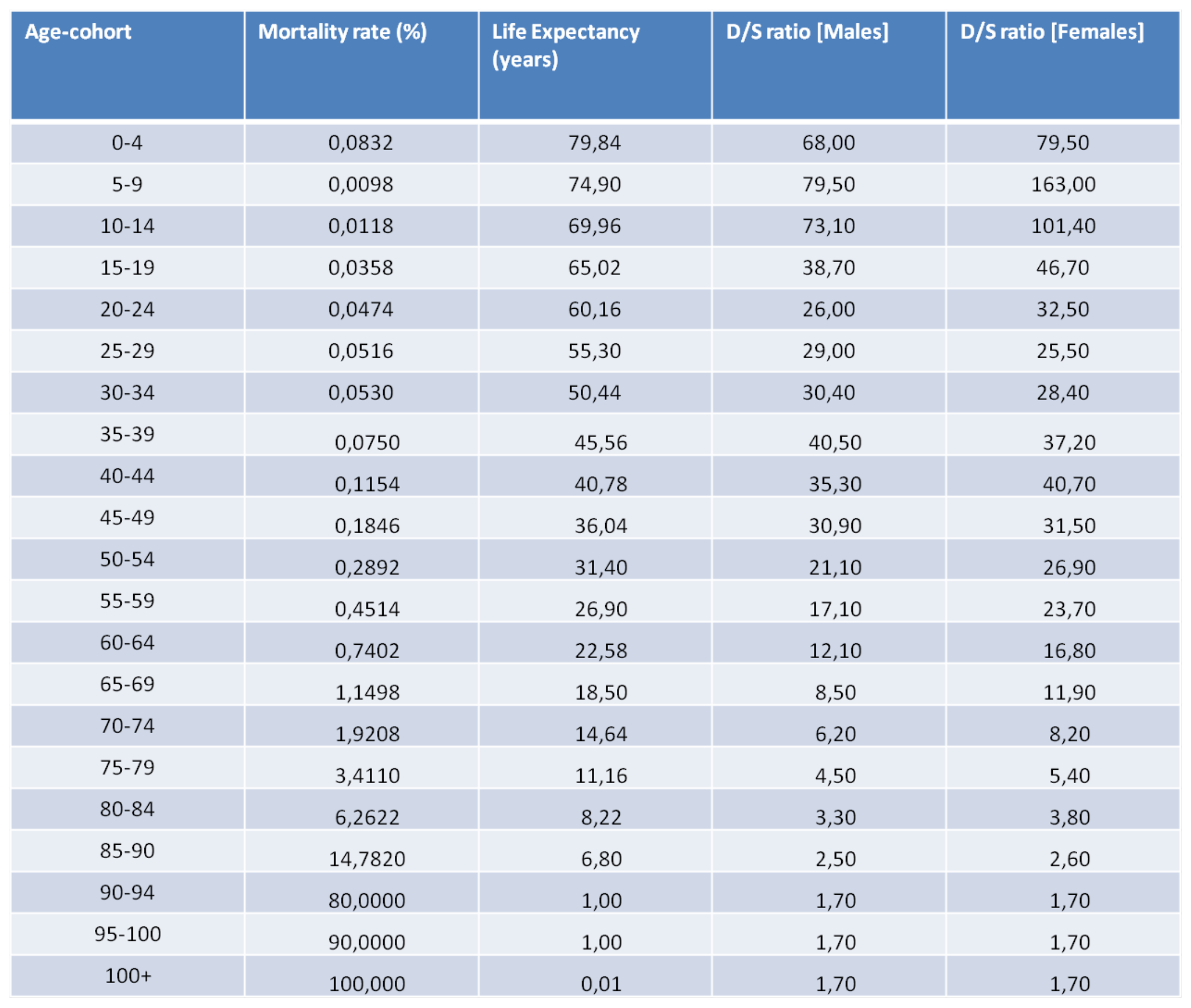}
\caption{Age-related Mortality rate, Expected Life and Deceased/Survivor ratio. \textit{Source: Mortality rate and life expectancy, Eurostat 2009. Decedent/Survivor ratio, mod. from \cite{RE_Przywara, Impact_Raitano}}.}
\label{fig: ReE_6}
\end{figure}
\begin{figure}[h!]
\centering
\includegraphics[scale=0.8]{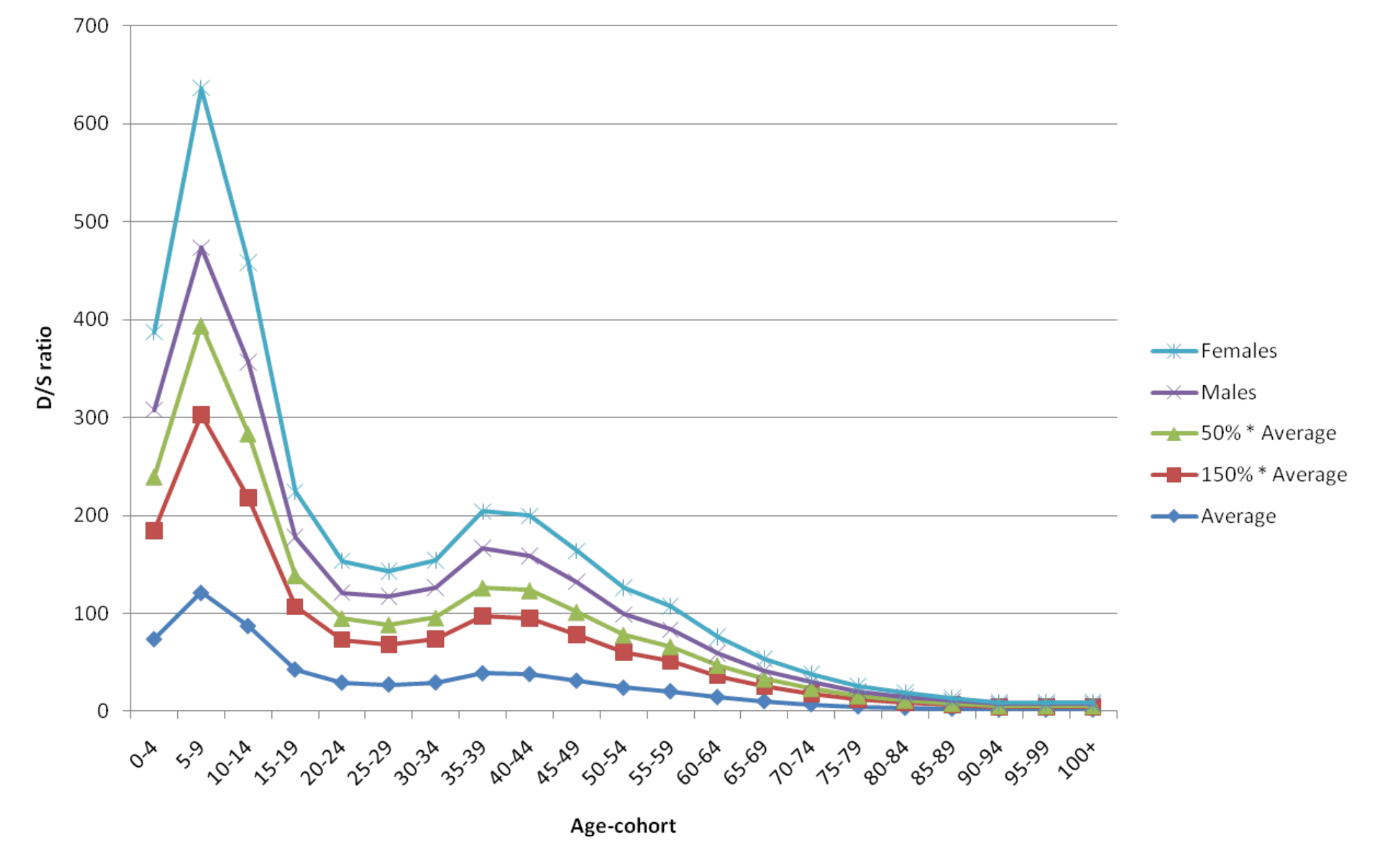}
\caption{Scenarios on Deceased/Survivor ratio. \textit{Source:  mod. from \cite{RE_Przywara}}.}
\label{fig: Impact_8}
\end{figure}

As far as spot age-related mortality rates \footnote{A set of mortality rates distinguished by age-cohort. Each one of them is applicable \textit{today} to persons within the same age-cohort.}, Expected Life and Deceased/Survivor ratio are concerned we referenced to \cite{RE_Przywara} (see fig. \ref{fig: ReE_6} and fig. \ref{fig: Impact_8}).


The general improvement of health reflects into longer expected life and lower mortality rates. 
Future expectations for mortality rates and life expectancy have been drawn starting from historical trends and adjusted according to a 3 months over year general health improvement \cite{RE_Przywara}.

\begin{figure}[h!]
\centering
\includegraphics[scale=0.8]{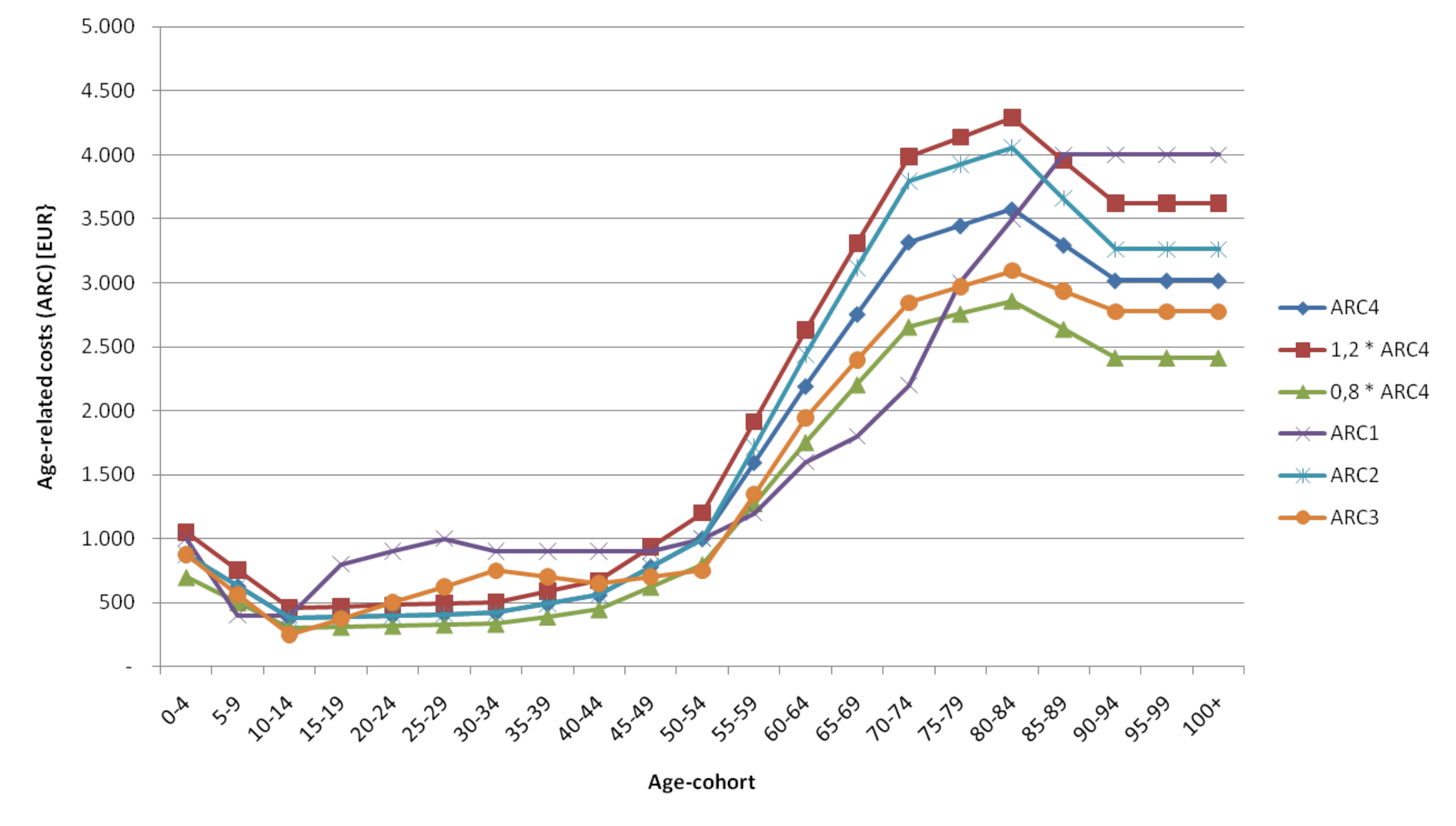}
\caption{Scenarios on age-related healthcare costs (ARC). \textit{Source: ARC1 mod. from \cite{Impact_EC}. ARC2, ARC3 and ARC4 mod. from \cite{RE_Przywara}}.}
\label{fig: ReE_7}
\end{figure}

Scenarios of age-related public healthcare costs, as shown in fig.\ref{fig: ReE_7}, have been adapted from various sources \cite{RE_Przywara, Impact_EC, Impact_Aprile}.


The impact of the crisis-related increased mortality (CRIMI) has been calculated assuming a functional relationship between healthcare expenditure (hcEXP) and the socio-economic-healthcare environment, i.e. a series of demographic ($d_1$, $d_2$, .., $d_M$) and healthcare related ($h_1$, $h_2$, .., $h_N$) parameters. For a given EM:
\begin{equation} \label{eq: Impact_5}
hcEXP = \Lambda^{[EM]}(d_1, d_2, .., d_M; h_1, h_2, .., h_N)
\end{equation}

\begin{figure}[h!]
\centering
\includegraphics[scale=0.5]{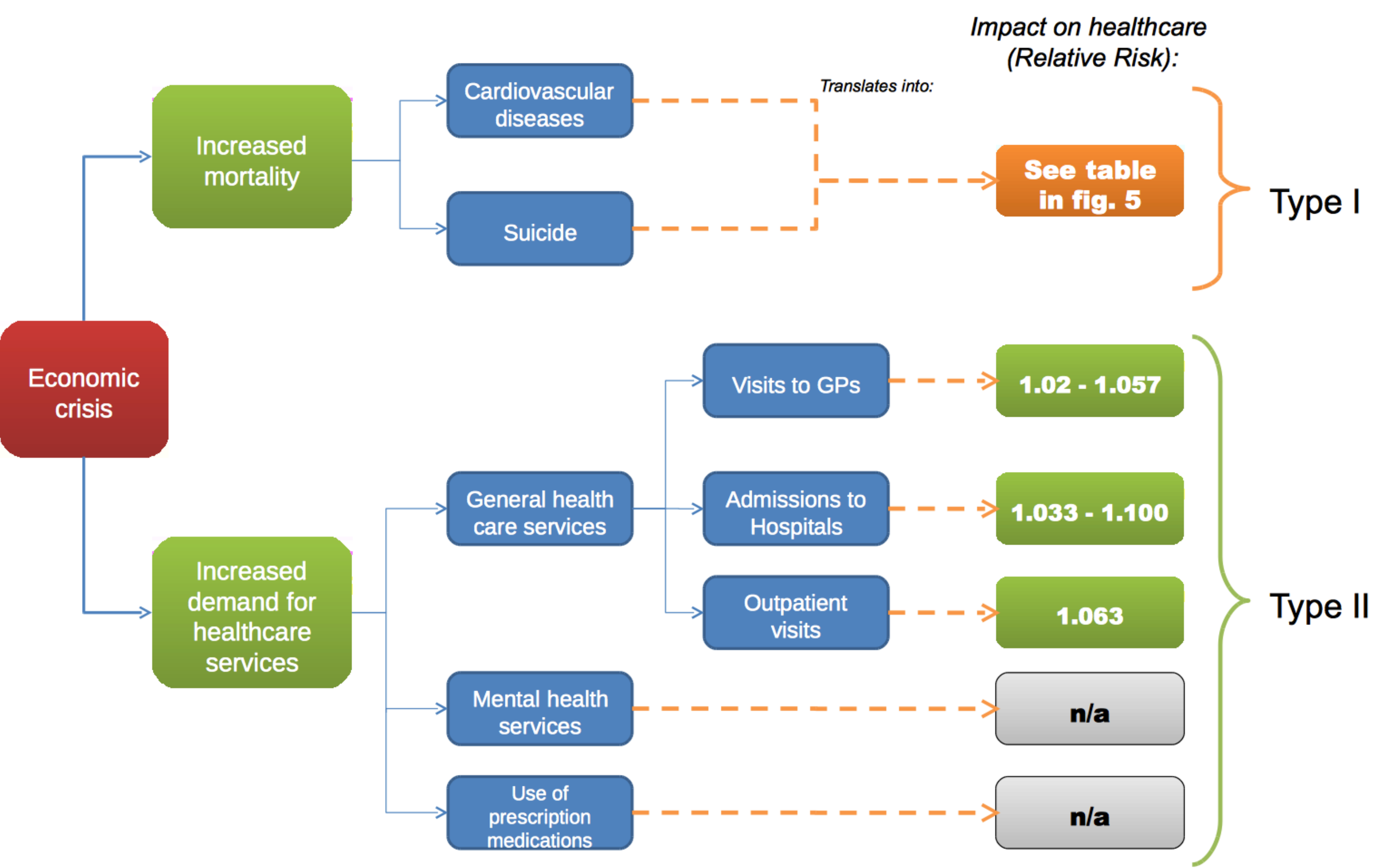}
\caption{Impact of the economic crisis on healthcare expenditure. The economic crisis causes (i) worsening of health conditions, that shows through an increased mortality associated to cardiovascular diseases and to suicides; (ii) an increased demand for healthcare services (more visits to GPs and specialists as well as increasing hospital admissions) that translates into higher costs for the NHS. The reader should note that while Type II effects \textit{directly} support an increased usage of healthcare services (hence, higher healthcare costs), Type I effects \textit{indirectly} translate into higher healthcare expenditure through worsening health conditions. \textit{n/a}: not assessed.}
\label{fig: ReE_1}
\end{figure}

\begin{figure}[h!]
\centering
\includegraphics[scale=0.5]{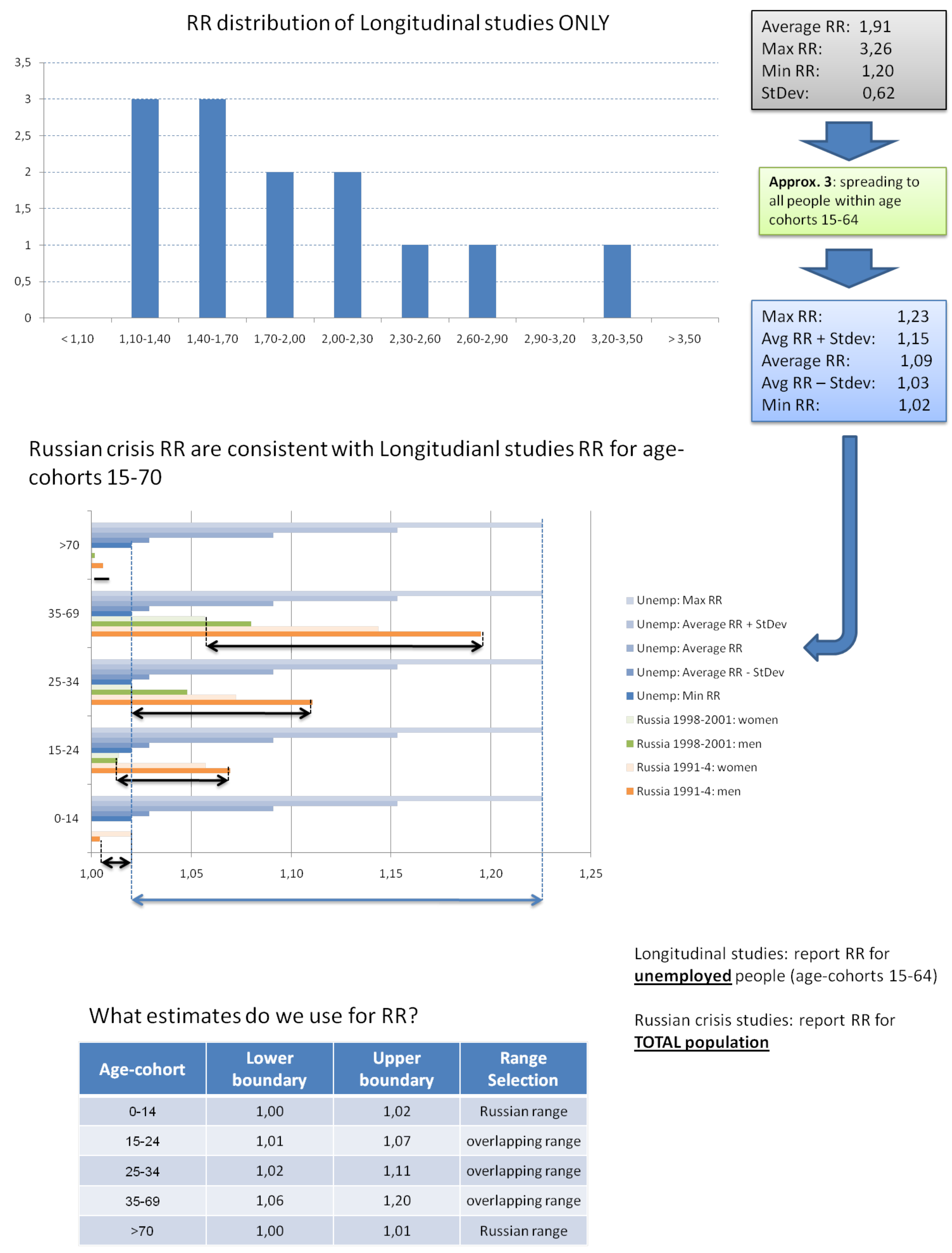}
\caption{\bb{(Top)} Distribution of RRs for \textit{increased mortality} (see fig.\ref{fig: ReE_1}). RRs have been taken from longitudinal cohort studies only as reported in fig. \ref{fig: Impact_3}. A basic statistical characterization of the distribution is reported in Box 1. By means of eq. \ref{eq: Impact_4} we have calculated the equivalent RR spread over the total population aged 15-64 (Box 2). \bb{(Middle)} Comparison between the RR variability inferred from the longitudinal cohort studies and those published for the first (1991-94) and second (1998-2001) economic downturns (see fig. \ref{fig: Impact_3}). The picture shows that the two different groups of studies overlap for middle age-cohorts. For middle age-cohorts, the overlapping ranges are selected. For age-cohorts with no overlap, the Russian levels have been selected. \bb{(Bottom)}. Values named \vv Lower and Upper boundary" will be used to estimate CRIMI.}
\label{fig: Impact_5}
\end{figure}

\noindent As previously discussed, the only two variables that we have used to parametrize the crisis impact on the public healthcare expenditure are (1) the increase in mortality rates and (2) the change in utilization of healthcare services.
Both effects are simplified as a rescaling of the corresponding base scenario according to the relative risk in fig. \ref{fig: ReE_1}.

Let's group all the parameters but the two above as $\mathcal{P}$. Let's call the mortality rates as $m^{(a)}_i$ where $a$ is the age group (e.g. 15-19 years) and $i$ is a date in the future (e.g. 2035) \footnote{The vector $m^{(a)}_{2010}$ corresponds to the second column in fig.\ref{fig: ReE_6}}. Be named $u^{(a)}$ the utilization rate at age cohort $a$.
Eq. \ref{eq: Impact_5} is rearranged as:
\begin{equation} \label{eq: Impact_7}
hcEXP = \Lambda^{[EM]}(m^{(a)}_i, u^{(a)};\mathcal{P})
\end{equation}
Within this framework, the CRIMI can be expressed as:
\begin{eqnarray} \label{eq_CRIMI}
CRIMI & = & \Lambda^{[EM]}(RR_{a, i}'^{MR}\cdot m^{(a)}_{i, BS}, u^{(a)}_{BS};\mathcal{P}_{BS}) |_{t=2015}  \nonumber \\
           &  - & \Lambda^{[EM]}(m^{(a)}_{i,BS}, u^{(a)}_{BS};\mathcal{P}_{BS}) |_{t=2015}
\end{eqnarray}
and the crisis-related increased utilization impact (CRIUI) as:
\begin{eqnarray} \label{eq_CRIUI}
CRIUI & = & \Lambda^{[EM]}(m^{(a)}_{i,BS}, RR_{a, i}^{UR}\cdot u^{(a)}_{BS};\mathcal{P}_{BS})|_{t=2015}  \nonumber \\
           &  - & \Lambda^{[EM]}(m^{(a)}_{i,BS}, u^{(a)}_{BS};\mathcal{P}_{BS})|_{t=2015}
\end{eqnarray}
where values with BS refer to the base scenario, i.e. the one without any crisis effect. $RR_{a, i}'^{MR}$ and $RR_{a}^{UR}$ are the relative risks that should be applied to mortality rates and, respectively, utilization rate (fig. \ref{fig: ReE_1}).

\begin{figure}[h!]
\centering
\includegraphics[scale=0.7]{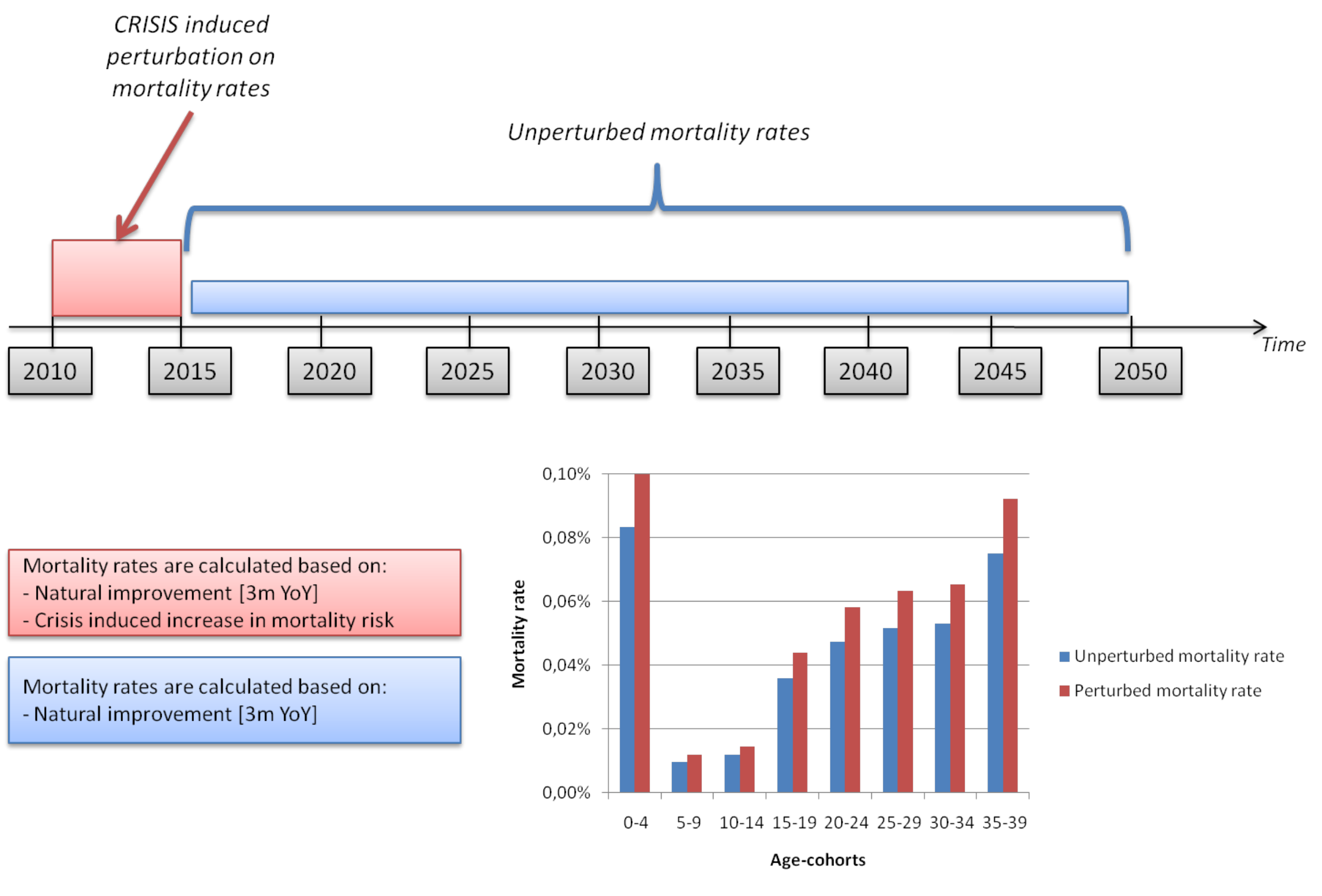}
\caption{CRIMI calculations: assumption on applied perturbation.}
\label{fig: Impact_7}
\end{figure}

\noindent Projections have been run over a discrete set of 5-years time values, hence the mortality rescaling $RR_{a, i}'^{MR}$ has been taken as (fig. \ref{fig: Impact_7}):
\begin{equation} \label{eq: Impact_6}
RR_{a, i}'^{MR} = 
\left \{
\begin{array} {rl}
RR_{a}'^{MR} & \mbox{, if  } t = 2015 \\
1                  & \mbox{, otherwise}
\end{array}
\right .
\end{equation}


\noindent As far as $RR_{a, i}^{UR}$ is concerned, all age-cohorts are assumed to have the same relative risk $RR^{UR}$: 
\begin{equation} \label{eq: Impact_UR}
RR_{a, i}'^{UR} = 
\left \{
\begin{array} {rl}
RR'^{UR} & \mbox{, if  } t = 2015 \\
1                  & \mbox{, otherwise}
\end{array}
\right .
\end{equation}

\noindent The total impact of the economic crisis on healthcare expenditure is calculated as:
\begin{eqnarray}
CRI = CRIMI + CRIUI
\end{eqnarray}

\section{Results}
%
\subsection{CRIMI estimates}
\begin{figure}[h!]
\centering
\includegraphics[scale=0.6]{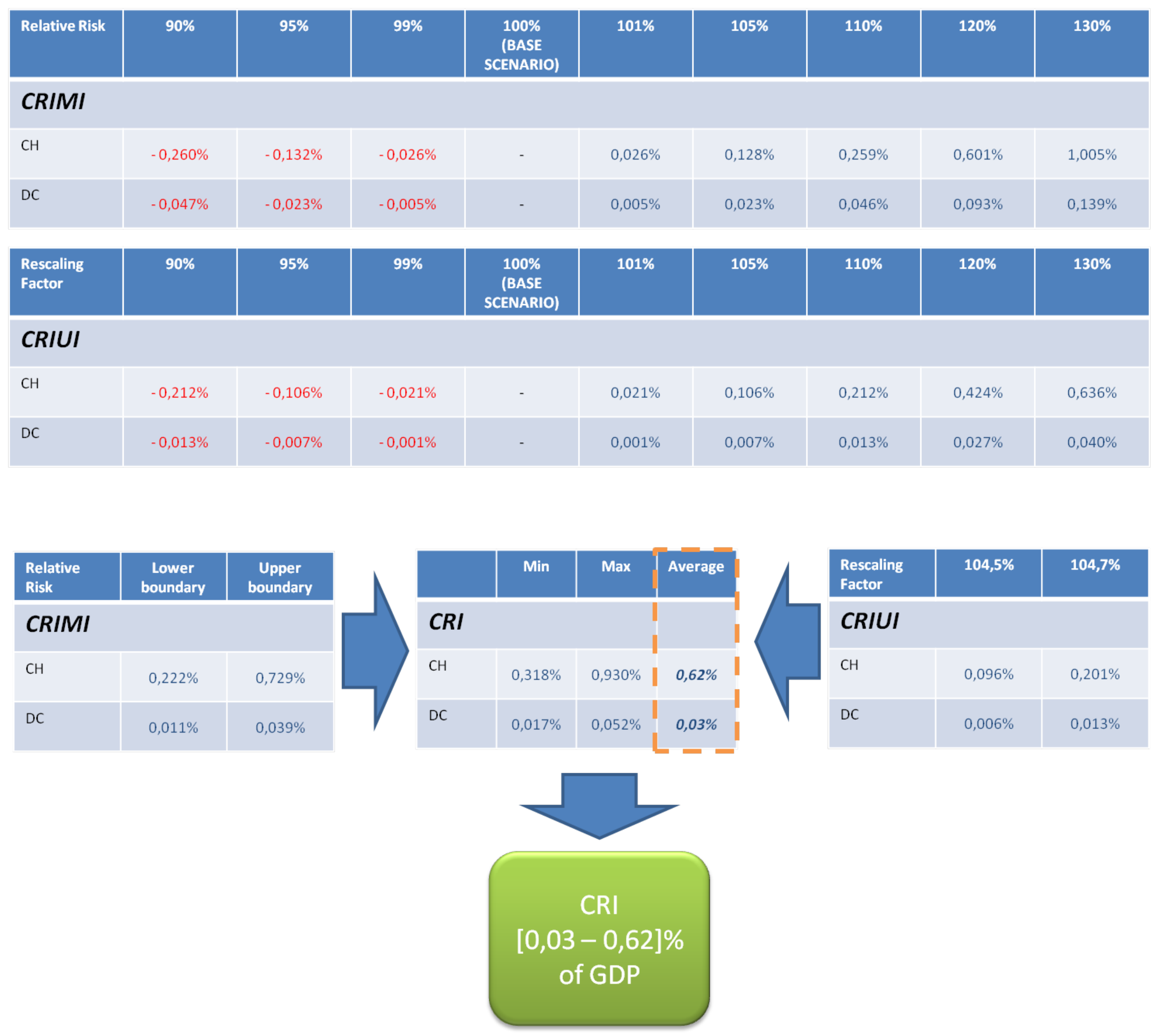}
\caption{Simulated CRIMI and CRIUI are shown for different econometric assumptions for a list of relative risks (RR) related to mortality rate and, respectively, healthcare utilization. The crisis effect is supposed to be taking place over a 5 years time, as pictured in fig. \ref{fig: Impact_7}. Please note that the results in the top table are not directly comparable to the ones showed in the CRIMI table on bottom left: the first ones are calculated by applying a RR which is \textit{uniform} throughout age-cohorts, while the second ones have been calculated based on \textit{age-dependent} RR tables reported in fig. \ref{fig: Impact_5}. CH: Constant Health, DC: Death-related Costs. Numbers in round brackets are negative numbers.}
\label{fig: ReE_10}
\end{figure}

Calculations for the impact of the economic crisis on 2015 healthcare public expenditure are reported in fig. \ref{fig: ReE_10}. In addition to calculating an estimate based on the impact of historical mortality and utilization rates, a sensitivity analysis to the estimates of mortality relative risk and utilization rate factor has been run. Estimates based on CH and DC models differ by an order of magnitude which is not surprising since mortality rates are treated differently by the two models.

\subsection{CRIUI estimates}
Once we have chosen (i) the EM, (ii) a reference scenario (BS) and (iii) a mapping between RR and mortality rates $m^{(a)}_{i,BS}$, the calculation of the impact of an increased relative risk of death is straightforward. In relation to the calculation of the impact of an increased relative risk of utilization rates, another assumption is needed: while mortality rates enter explicitly within the dynamics of future healthcare costs, the same does not hold for the utilization rate of healthcare services and a model assumption becomes essential.\footnote{Within the CH model, the improvement in the age-related healthcare costs is a function of the change in the mortality rates; within the DC model, the healthcare expenditure depends on the number of expected deaths.} 

\subsubsection*{Assumption n.3}
By definition of age-related healthcare costs pro-capita $c_i^{(a)}$, eq. \ref{eq: Impact_5} can be written as:
\begin{equation}
hcEXP = \sum_a N_i^{(a)} \cdot c_i^{(a)}
\end{equation}
where $N_i^{(a)}$ is the number of persons falling within age-cohort $a$.
Total healthcare expenditure can be divided into four typologies: (i) hospital (i.e. in-patients), \textit{H}; (ii) pharmaceutical, \textit{P}; (iii) specialistic (i.e. out-patients), \textit{S}; (iv) general practitioners, \textit{GP}; (v) rehabilitation, \textit{R}; (vi) other minors, \textit{m} \footnote{Definitions for $H, P, S, GP, R$ and $m$ available at 2008 Italian Treasury economic statement, \textit{www.salute.gov.it}.}. Thus we can also write:
\begin{equation}
hcEXP = H + P + S + GP +  R + m
\end{equation}
Data available from epidemiological studies show that each one of these cost item is increased as a consequence of unemployment by its own RR. Unfortunately epidemiological RR are available for $H$, $P$ and $S$ only (see fig. \ref{fig: ReE_1}) \footnote{An RR of 1.00 has been assumed for P, R and m.}. Nevertheless this should not cause any major concern since $R$ and $m$ represent a minor amount of the Italian public healthcare expenditure.
The rescaling factor (RF) that should be applied to the age-related costs in order to model the increase in healthcare utilization rates can be calculated as:
\begin{equation} \label{eq: Impact_8}
RF = RR_{H} \cdot \%H + RR_{P} \cdot  \%P + RR_{S} \cdot \%S + RR_{GP} \cdot \%GP + RR_{R} \cdot \%R + RR_{m} \cdot \%m
\end{equation}
Based on Italian Treasury 2008 data, we have \%H=71\%, \%P=10\%, \%S=4\%, \%GP=6\%, \%R=2\% and \%m=7\%. Accordingly, RF falls within the range 1.045 - 1.095. CRIUI is thus calculated as: 
\begin{eqnarray}\label{eq: Impact_9}
CRIUI & = & \Lambda^{[EM]}(m^{(a)}_{i, BS}, RF \cdot c^{(a)}_{BS};\mathcal{P}_{BS}) |_{t=2015}  \nonumber \\
           &  - & \Lambda^{[EM]}(m^{(a)}_{i,BS}, c^{(a)}_{BS};\mathcal{P}_{BS}) |_{t=2015}
\end{eqnarray}

\begin{figure}[h!]
\centering
\includegraphics[scale=0.7]{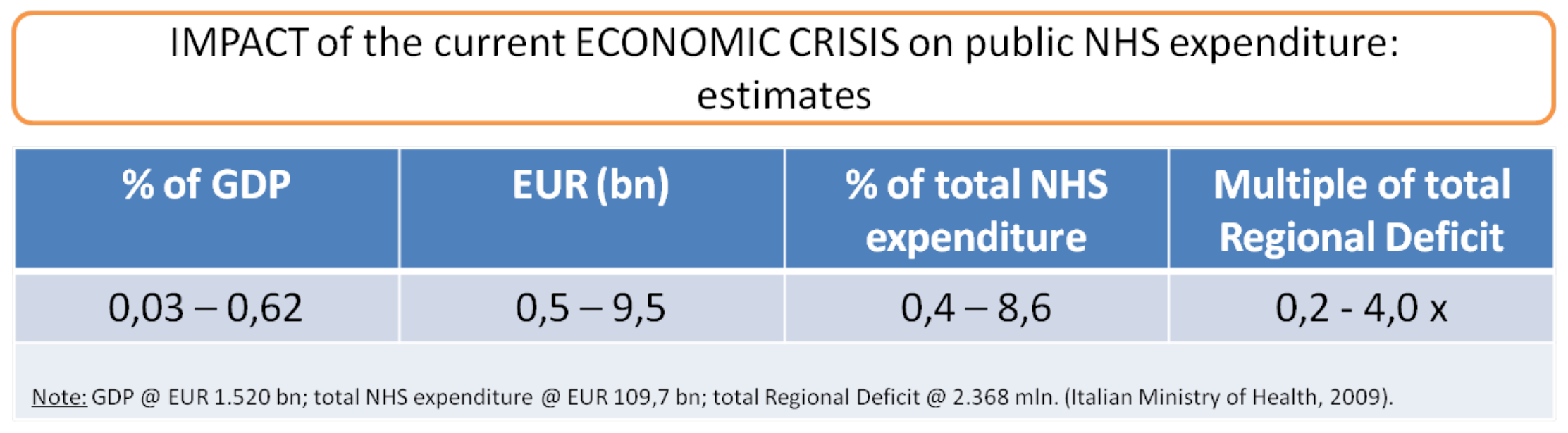}
\caption{Estimated impact of current economic crisis on public NHS healthcare expenditure for year 2015.}
\label{fig: Impact_9}
\end{figure}

\vskip 5mm
\noindent and results are reported in fig. \ref{fig: ReE_10}. By combining the ranges calculated for CRIMI and CRIUI we get as \bb{estimate for the total crisis-related impact on public healthcare expenditure}, the range \bb{0.03 - 0.62 \%} of Italian GDP.
Using GDP level for 2009 \footnote{GDP: EUR 1,520,346 mln; total NHS expenditure: EUR 109.7 bn; total regional Deficit: EUR 2,368 mln. Source: ISTAT, year 2009.} we get an estimate for CRI: \bb{EUR 0.5 - 9.5 bn} (fig. \ref{fig: Impact_9}).

\section{Limits} 
\label{sec: Limits}
%
\noindent
To the best of our knowledge there are no conclusive and comprehensive studies in the scientific literature on the effect of an economic crisis in terms of mortality rates and utilization rates of healthcare services. The epidemiological studies that we reported show an ample range of variability when the same parameters are considered. Moreover the effects of an economic crisis are highly dependent on the type of healthcare system in place. As a general example one could compare the totally opposite results that has been found for two countries involved in the same East Asian crisis, i.e. Indonesia and Thailand \cite{Impact_Waters}.

\noindent Our estimates have privileged epidemiological studies related to European countries and, among these, we focused on those healthcare systems similar to the Italian one. 

\noindent If on the mortality relative risk we have been able to average on a set of ten studies, much less information was available for the utilization rates of healthcare services. As a consequence the estimates on CRIUI have a higher uncertainty than those on CRIMI.

Estimated model parameters could be a weak point of our calculations. A wider set of data would need to be collected from other epidemiological studies. Unfortunately this is a difficult task since there is no database reporting the sought-after effects. Additional complications come from the lack of standardization among epidemiological studies in terms of reported parameters.

Two different approaches (CH and DC) have been used to assess the impact of the crisis on potential public healthcare expenditure. None of them is deemed to be perfect or superior, but each offers some insight into this difficult issue.
There is no evidence at all that one of the two is preferable to the other and indeed they are both taken into account when elaborating demographic projections at the European Community level \cite{RE_Przywara}.
\noindent Further insight on the actual dynamics between unemployment, mortality and usage of healthcare services could help improve our results.

\section{Conclusions}    
\label{Conclusions}
%
\noindent
One of the main issues affecting the Italian NHS is the healthcare deficit: according to current agreements between the Italian State and its Regions \cite{ConferenzaStatoRegioni} public funding of regional NHS is now limited to the amount of regional deficit and is subject to previous assessment of strict adherence to constraint on regional healthcare balance-sheet. Many Regions with previously uncontrolled healthcare deficit have now to plan their \vv Piano di Rientro" (PdR) and submit it for the approval of the Italian Ministry of Economy and Finance. 
Those Regions that will fail to comply to deficit constraints will suffer cuts on their public NHS financing.   

The estimated impact of the economic crisis is comparable to the healthcare deficit of Italian Regions (EUR 3-5 bn), hence any serious Health Planning over the next five years needs to take the problem of possible negative financial implications of current economic crisis into account.

There is much that can be done to mitigate the impact of the financial crisis. Clearly, even in times of crisis, the first goal of the Italian NHS should be to guarantee that health is protected. A smart Health Planning can make sure health-spending is managed appropriately. Italy should recognize current crisis as an opportunity to undertake financial and sectoral reforms. It should pursue a multi-sectorial response, with an emphasis in the health sector. Although short-term measures to mitigate negative consequences are urgent, a longer term perspective aimed at making the Italian health sector more resilient in the future is needed.

\newpage

\section{Figures}
\label{sec:Figures}
%

\cleardoublepage
\bibliographystyle{JHEP}

\begin{thebibliography}{99}


\bibitem{IMFCountryReport} 
  ``Changes in morbidity and medical care utilization after the recent economic crisis in the Republic of Korea."
 International Monetary Fund October 2010 Country Report.\  {\bf } (2010).

\bibitem{Fachin} 
S.~Fachin and A.~Gavosto,
  ``The decline in Italian Productivity: a study in the estimation of long-run trends in total factor productivity with panel cointegration methods."
LLEE Working Document \ {\bf 50}, (2007)

\bibitem{Intro_Bassanetti} 
A.~Bassanetti, M.~Iommi, C.~Jona-Lasinio, and F.~Zolin,
  ``La crescita dell'economia italiana negli anni novanta tra ritardo tecnologico e rallentamento della produtivit\`a."
 Temi di discussione della Banca d'Italia, n. 539.\  {\bf } (2004).

\bibitem{HealthEconStatus1}
J.~Adda, T.~Chandola and M.~Marmot,
  ``Socio-economic status and health: causality and pathways."
Journal of Econometrics {\bf 112}, 57-63 (2003)

\bibitem{HealthEconStatus2}
D.~E.~Bloom, D.~Canning and J.~Sevilla,
  ``The effect of health on economic growth: a production function approach."
World Development {\bf 32}, 1-13 (2004)

\bibitem{HealthEconStatus3}
J.~Bound, M.~Schoenbaum, T.~R.~Stinebrickner and T.~Waidmann,
  ``The dynamic effects of health on the labour force transitions of older workers."
Labour Economics, 179-202 (1999)

\bibitem{HealthEconStatus4}
P.~Contoyannis and A.~M.~Jones,
  ``Socio-economic status, health and lifestyle."
Journal of Health Economics, {\bf 23} 965-995 (2004)

\bibitem{HealthEconStatus5}
J.~Currie and R.~Hyson,
  ``Is the impact of health shocks cushioned by socioeconomic status? The case of low birthweight."
NBER Working Paper (1999)

\bibitem{HealthEconStatus6}
M.~A.~Winkleby, D.~E.~Jatulis, E.~Frank and S.~P.~Fortmann,
  ``Socioeconomic status and health: how education, income and occupation contribute to risk factors for cardiovascular disease."
American Journal of Public Health, {\bf 82} 816-820 (1992)

\bibitem{HealthEconStatus7}
M.~G.~Marmot, M.~Kogevinas and M.~A.~Elston,
  ``Social-economic status and disease."
Annual Review of Public Health, {\bf 8} 111-135 (1987)

\bibitem{HealthEconStatus8}
N.~Adler, T.~Boyce and M.~A.~Chesney,
  ``Socioeconomic status and health."
American Psychologist, {\bf } 15-22 (1994)

\bibitem{PastCrisisConseq1}
H.~Kim, W.~J.~Chung and Y.~J.~Song,
  ``Changes in morbidity and medical care utilization after the recent economic crisis in the Republic of Korea."
Bulletin of the World Health Organization, {\bf 81} 8 (2003)

\bibitem{SubpCrisisConseq1}
S.~Kwon and Y.~Jung,
  ``The impact of the global recession on the health of the people in Asia."
3rd China-ASEAN Forum on Social Development and Poverty Reduction, {\bf } (2009)

\bibitem{SubpCrisisConseq2}
M.~Landler,
  ``Financial Chill May Hit Developing Countries."
New York Times, September 25, {\bf } (2008)

\bibitem{SubpCrisisConseq3}
D.~Roodman,
  ``History says financial crisis will suppress aid."
Center for Global Development, 13 Oct 2008. {\bf } (2008)




\bibitem{Polder}
For an introduction to Deceased/Survivor Ratio and Death-related costs see:\\
A.~Wong, P.~H.~M.~Van Baal, H.~C.vBoshuizen and J.~J.~Polder,
  ``Exploring the influence of proximity to death on disease-specific hospital expenditures: a carpaccio of red herrings."
Health Economics {\bf 20}(4) 379-400 (2011);
A.~Wong,
``Describing, explaining and predicting health care expenditures with statistical methods.". 
\href{http://arno.uvt.nl/show.cgi?fid=122354}{[PhD thesis at Tilburg University]};
M.~Raitano,
``The Impact of Death-Related Costs on Health Care Expenditure: A Survey" {\bf } (2006);
S.~Gabriele, C.~Cislaghi, F.~Costantini, F.~Innocenti, V.~Lepore, F.~Tediosi, M.~Valerio and C.~Zocchetti,
``Demographic factors and health expenditure profiles by age: the case of Italy"
\href{http://www.enepri.org/files/AHEAD/Reports/}{[ENEPRI Report WP7]};



\bibitem{EP1}
M.~Seshamani and A.~Gray,
  ``Time to death and health expenditure: an improved model for the impact of demographic change on health care costs."
Age and Ageing {\bf 33} 556-561 (2004)

\bibitem{EP2}
T.~T.~Dang, P.~Antolin and H.~Oxley,
Economics Department Working Papers No. 305. Organization for Economic Co- operation and Development {\bf } (2001)

\bibitem{EP3}
T.~Miller,
  ``Increasing longevity and Medicare expenditures."
Demography {\bf 38} 215-226 (2001)

\bibitem{RE_Przywara}
B.~Przywara,
  ``Projecting future health care expenditure at European level: drivers, methodology and main result."
EUROPEAN ECONOMY, Economic Papers {\bf 417} (2010)




\bibitem{Impact_Men}
T.~Men, P.~Brennan, P.~Boffetta and D.~Zaridze,
  ``Russian mortality trends for 1991-2001: analysis by cause and region."
BMJ {\bf 327}, 964 (2003)

\bibitem{Impact_Sachs}
J.~Sachs,
  ``Understanding \vv shock therapy"."
Occasional paper. London: Social Market Foundation {\bf } (1994)

\bibitem{Impact_Wedel}
J.~Wedel,
  ``Collision and Collusion: The strange case of Western aid to Eastern Europe."
New York: St. Martin's {\bf } (2001)

\bibitem{Impact_Murrell}
P.~Murrell,
  ``What is shock-therapy? What did it do in Poland and Russia?."
Post-Soviet Affairs {\bf 9}(2) 111-140 (1993)

\bibitem{Impact_Kontorovich}
V.~Kontorovich,
  ``The Russian health crisis and the economy."
Communist and Post-Communist Studies {\bf 34} 221-240 (2001)

\bibitem{Impact_Kim}
WHO,
  ``Changes in morbidity and medical care utilization after the recent economic crisis in the Republic of Korea."
Bulletin of the World Health Organization {\bf 81} (2003)

\bibitem{Impact_Khang}
Y.~Khang, J.~W.~Lynch and G.~A.~Kaplan,
  ``Impact of economic crisis on cause-specific mortality in South Korea."
International Journal Epidemiology {\bf 34} 1291-1301 (2005)

\bibitem{Impact_Waters}
H.~Waters, F.~Saadah and M.~Pradhan,
  ``The impact of the 1997 East Asian economic crisis on health and health care in Indonesia."
Health policy and planning {\bf 18} 172-181 (2003)

\bibitem{Moreno}
R.~Moreno, G.~Pasadilla and E.~Remolona,
  ``Asia's financial crisis: lessons and policy responses."
Pacific Basin Working Paper Series {\bf }PB98-02 (1998)

\bibitem{Intereconomics}
G.~Aschinger,
  ``An Economic Analysis of the East Asia Crisis."
Pacific Basin Working Paper Series {\bf }March/April 55-63 (1998)

\bibitem{Impact_Siddique}
C.~M.~Siddique and C.~D'Arcy,
  ``Unemployment and health: an analysis of the Canada health survey."
International Journal Health Services {\bf 15} 609-635 (1985)

\bibitem{Impact_Moser1}
D.~Jones, K.~Moser and A.~Fox,
  ``Unemployment and mortality in the OPCS longitudinal study."
The Lancet {\bf 2} 8415, 1324-1329 (1984)

\bibitem{Impact_Moser2}
K.~Moser, A.~Fox, D.~Jones and P.~Goldblatt,
  ``Unemployment and mortality: further evidence from the OPCS longitudinal study 1971-81."
The Lancet {\bf 327} 8477, 365-367 (1986)

\bibitem{Impact_Beale1}
N.~Beale and S.~Nethercott,
  ``The nature of unemployment morbidity. Recognition."
Journal of the Royal College of General Practitioners {\bf 38} 310, 197-199 (1988)

\bibitem{Impact_Beale2}
N.~Beale and S.~Nethercott,
  ``The nature of unemployment morbidity. Description."
Journal of the Royal College of General Practitioners {\bf 38} 310, 200-202 (1988)

\bibitem{Impact_Yuen}
P.~Yuen and R.~Balarajan,
  ``Unemployment and patterns of consultation with the general practitioner."
BMJ {\bf 298} 1212-1214 (1989)

\bibitem{Impact_Martikainen}
P.~Martikainen,
  ``Unemployment and mortality among Finnish men, 1981-1985."
BMJ {\bf 301} 407-411 (1990)

\bibitem{Impact_Iversen}
L.~Iversen, O.~Andersen and P.~Andersen,
  ``Unemployment and mortality in Denmark, 1970-80."
BMJ {\bf 295} 879-884 (1987)

\bibitem{Impact_Costa}
C.~Costa and N.~Segman,
  ``Unemployment and mortality."
BMJ {\bf 294} 1550-1551 (1987)

\bibitem{Impact_Linn}
M.~W.~Linn, R.~S.~Sandifer and S.~Stein,
  ``Effects of unemployment on mental and physical health."
Am J Public Health {\bf 75} 502-506 (1985)

\bibitem{Impact_Jin}
R.~L.~Jin, C.~P.~Shah, T.~J.~Svoboda,
  ``The impact of unemployment on health: a review of the evidence."
Canadian Medical Association Journal {\bf 153}(5) 529-540 (1985)

\bibitem{Impact_Frey}
J.~Frey,
  ``Unemployment and health in the US."
BMJ {\bf 284} 1112 (1982)


\bibitem{Impact_EC}
European Commission,
Special Report n.4/2005 - European Economy (Directorate-General for Economic and Financial Affairs).

\bibitem{Impact_Aprile}
R.~Aprile,
  ``The impact of ageing on health and long-term care: the case of Italy."
Bank of Italy, Workshop on Fiscal Sustainability: Analytical Developments and Emerging Policy Issues. {\bf } (2008)
\href{http://www.bancaditalia.it/studiricerche/convegni/atti/fiscal_sustainability/session_3/Aprile.pdf}{[The impact of ageing on health and long-term care: the case of Italy.]}

\bibitem{Impact_Raitano}
M.~Raitano,
ENEPRI Research Report N. 17 {\bf } (2006)
\href{http://www.enepri.org}{[ENEPRI Research Report N. 17]}


\bibitem{ConferenzaStatoRegioni}
ConferenzaStato-Regioni,
Rep. n. 243/CSR del 3 dicembre 2009
\href{http://www.governo.it/GovernoInforma/Dossier/patto salute/}{[http://www.governo.it/GovernoInforma/Dossier/patto salute/]}


\end{thebibliography}


\end{document}